\documentstyle[12pt]{article}
\def\thebibliography#1{\section*{\normalsize\underline{References}\markboth
  {References}{References}}\list
  {[\arabic{enumi}]}{\settowidth\labelwidth{[#1]}
    \leftmargin\labelwidth
    \advance\leftmargin\labelsep
    \usecounter{enumi}}
    \def\newblock{\hskip .11em plus .33em minus -.07em}
    \sloppy
    \sfcode`\.=1000\relax}

\def\ackno{\section*{\normalsize\bf ACKNOWLEDGMENTS}}

\input psfig

\begin{document}


\renewcommand{\thetable}{\Roman{table}}

\begin{center}
FIRST-PRINCIPLES CALCULATION OF THE STRUCTURE OF MERCURY
\end{center}

\noindent MICHAEL J. MEHL

\noindent Complex Systems Theory Branch, Naval Research
Laboratory, Washington, DC 20375-5345\\[0.125in]

\noindent
{\bf ABSTRACT}: Mercury has perhaps the strangest behavior of any of
the metals.  Although the other metals in column IIB have an $hcp$
ground state, mercury's ground state is the body centered tetragonal
$\beta$Hg phase.  The most common phase of mercury is the
rhombohedral $\alpha$Hg phase, which is stable from 79K to the
melting point and meta-stable below 79K.  Another rhombohedral phase,
$\gamma$Hg, is believed to exist at low temperatures.
First-principles calculations are used to study the energetics of
the various phases of mercury.  Even when partial spin-orbit effects are
included, the calculations indicate that the hexagonal close packed
structure is the ground state.  It is suggested that a better
treatment of the spin-orbit interaction might alter this result.

\section*{\normalsize\bf INTRODUCTION}
\label{sec:intro}
{~}\\[-0.5in]

Of all the metals in the periodic table, mercury has the most
interesting behavior.  A liquid at room temperature, the metal only
solidifies below 234K.  This phase, denoted $\alpha$Hg, (Pearson
symbol hR1, space group $R\overline{6}m$, {\em Strukturbericht}
designation $A10$), is a rhombohedral structure with one atom per
unit cell and the primitive vectors at an angle of
70$^\circ$44.6'\cite{wycoff63}.  Although $\alpha$Hg has been seen
experimentally down to 5K, below 79K the ground state is the phase
$\beta$Hg (prototype $\alpha$Pa, Pearson symbol tI2, space group
I4/mmm, {\em Strukturbericht} designation A$_a$), a body-centered
tetragonal ($bct$) phase with a $c/a$ ratio of
0.7071\cite{donohue74}.  Nearer to room temperature, $\alpha$Hg can
be transformed to $\beta$Hg by applying pressure.  A third,
meta-stable form, $\gamma$Hg, has also been
observed\cite{donohue74}.  Like $\alpha$Hg it is rhombohedral with
one atom per unit cell, but the primitive vectors are at an angle of
about 50$^\circ$.

Since these phases of mercury all involve only one atom per unit
cell, first-principles {\em ab initio} methods are relatively easy
to use.  An extensive literature search, however, found very few
studies of mercury\cite{ballone89,neisler87,psingh94}, with only one
\cite{psingh94} including relativistic effects, which are important
for all of the late fifth-row elements, and that only for metallic
clusters.  It is useful, therefore, to perform a series of
first-principles calculations for the various phases of mercury.
This paper presents results using the full potential, Linearized
Augmented Plane Wave (LAPW) method\cite{andersen75,wei85,singh94}
using the Hedin-Lundqvist\cite{hedin71} parametrization of the Local
Density Approximation (LDA)\cite{kohn65} to Density Functional
Theory (DFT)\cite{kohn65,hohenberg64}.  The calculations were
initially performed in the scalar-relativistic
approximation\cite{koelling75}, which essentially ignores the
spin-orbit interaction while maintaining the remaining relativistic
contributions.  Spin-orbit corrections were then included using the
``second-variational'' method\cite{macdonald80}.

\section*{\normalsize\bf THE STRUCTURES OF MERCURY}
\label{sec:struc}
{~}\\[-0.5in]

\begin{figure}[htb]
\centerline{\psfig{file=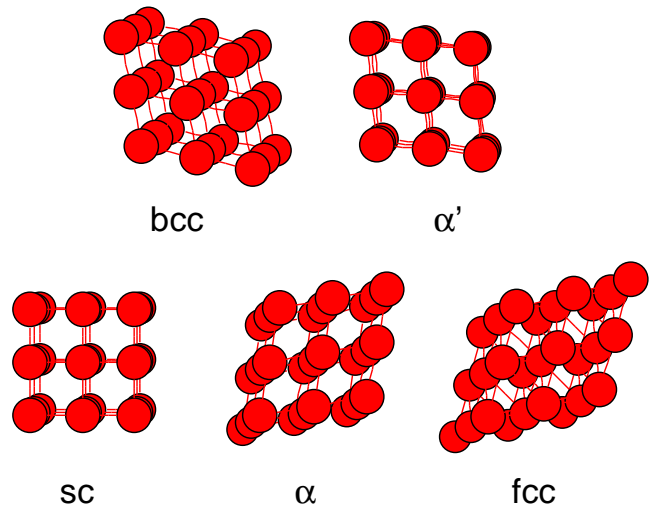,height=3.0in}}
\caption{Several of the structures found in the rhombohedral
system\protect{\ref{equ:rhom}}.  The $\alpha'$ phase shown here has
an angle $\alpha$ between $\pi/2$ and $\cos^{-1} (-1/3)$.  The
$\gamma$Hg phase would be to the right of the $fcc$ phase in this
plot.}
\label{fig:alpha}
\end{figure}

The primary structures of mercury are the rhombohedral $\alpha$Hg
phase and the body-centered tetragonal $\beta$Hg phase.  Each phase
can be described by two parameters: the volume and a parameter
describing the orientation of the primitive vectors.  In $\alpha$Hg
this parameter is the angle $\alpha$ between the primitive vectors.
In $\beta$Hg the parameter is the $c/a$ ratio of the tetragonal unit
cell.  Special values of these parameters lead to higher symmetry
unit cells.  The primitive vectors of the rhombohedral $\alpha$Hg
phase can be written in the form
\begin{equation}
\begin{array}{ccccccccccc}
{\bf a}_1 & = & a & ( & 1 + x & , & x & , & x & ) & \\
{\bf a}_2 & = & a & ( & x & , & 1 + x & , & x & ) & , ~ x = \frac13 (
 \sqrt{(1 + 2 \cos \alpha)/(1 - \cos \alpha)} - 1) \\
{\bf a}_3 & = & a & ( & x & , & x & , & 1 + x & ) & ~ ,
\end{array}
\label{equ:rhom}
\end{equation}
where $\alpha$ is the angle between the primitive vectors.  There
are several special values of this angle.  At $\alpha = 0$ the
vectors (\ref{equ:rhom}) are collinear, while at $\alpha = 2 \pi/3$
they are coplanar.  These unphysical situations bound the range of
$\alpha$.  Several high symmetry lattices can also be obtained from
(\ref{equ:rhom}).  At $\alpha = \pi/3$ we find the $fcc$ lattice, at
$\alpha = \pi/2$ the simple cubic ($sc$) lattice, and at $\cos\alpha
= -1/3$ the $bcc$ lattice.  Because of these symmetries, a plot of
the energy $E(V,\alpha)$ at fixed volume $V$ would show the energy
diverging as $\alpha$ approached both zero and $2 \pi/3$, with
extremal points at $\alpha = \pi/3$, $\pi/2$, and $\cos^{-1}
(-1/3)$.  Since the $\alpha$Hg phase has $\alpha \approx 70^\circ$
degrees, this phase will appear between the $fcc$ and $sc$ phases.
The $\gamma$Hg phase, with $\alpha \approx 50^\circ$, has a smaller
angle than the $fcc$ phase.  Figure~\ref{fig:alpha} shows several of
the phases found in the rhombohedral system.

\begin{figure}[htb]
\centerline{\psfig{file=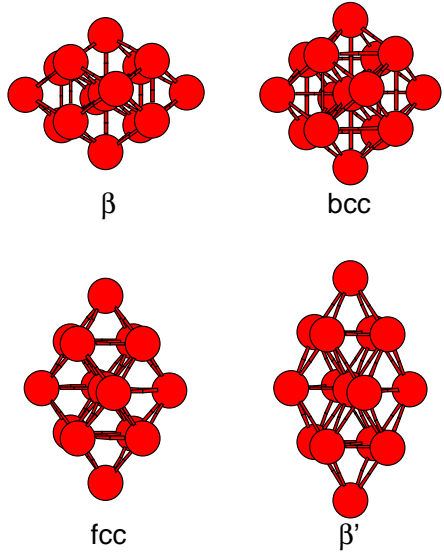,height=3.0in}}
\caption{Several of the structures found in the $bct$
system\protect{\ref{equ:bct}}.  The $\beta'$ phase shown here has
a value of $c/a$ between the $fcc$ and $bcc$ phases.}
\label{fig:beta}
\end{figure}

The primitive vectors of the $bct$ $\beta$Hg phase can be written in
the form
\begin{equation}
\begin{array}{ccccccccc}
{\bf a}_1 & = & ( & a & , & 0 & , & 0 & ) \\
{\bf a}_2 & = & ( & 0 & , & a & , & 0 & ) \\
{\bf a}_3 & = & ( & \frac12 a & , & \frac12 a & , & \frac12 c & ) ~
{}.
\end{array}
\label{equ:bct}
\end{equation}
This lattice is identical to the $fcc$ lattice when $c/a =
\sqrt{2}$, and to the $bcc$ lattice when $c/a = 1$.  It is
interesting to note that, within experimental error, the $\beta$Hg
lattice has $c/a = 1/\sqrt{2}$.  At this value of $c/a$ each mercury
atom has two nearest neighbors, located directly above and below the
atom along the $z$ axis, at $\pm (2 {\bf a}_3 -{\bf a}_1 - {\bf
a}_2)$, and eight next-nearest neighbors.  The mercury atoms thus
form chains running along the $z$ direction.  Figure~\ref{fig:beta}
shows several phases in the $bct$ system.

\section*{\normalsize\bf FIRST-PRINCIPLES CALCULATIONS}
\label{sec:scalar}
{~}\\[-0.5in]

The first set of calculations were performed using the LAPW method
in the ``scalar-relativistic'' approximation outlined above.  This
method essentially ignores the spin-orbit interaction, keeping the
remainder of the relativistic corrections.  In all of the
calculations the muffin-tin radius was set at $R_{MT} = 2.2$ atomic
units, a size chosen to allow large strains of the form
(\ref{equ:rhom}) or (\ref{equ:bct}) while keeping the muffin tins
{}from touching.  The momentum cutoff $K_{max}$ was chosen so that
$R_{MT} K_{max} = 10.5$, yielding typical secular-equation
dimensions of about $300 \times 300$.  Increasing the cutoff to 11.5
decreases the energy by about 0.3m~Ry for all structures and
volumes.  The K-point meshes were chosen using a regular mesh evenly
spaced along the primitive vectors.  Meshes of 150-200 K-points in
the irreducible Brillouin zone, depending on the structure, yield
total energies accurate to about 0.3~mRy compared to larger K-point
meshes.  The energies computed here are thus accurate to about
0.5~mRy.

Comparing the scalar-relativis\-tic and spin-orbit energies obtained
{}from a Liberman-based atomic code\cite{liberman71} show that the
spin-orbit interaction contributes 29.5 mRy to the total energy.
This is a relatively large contribution, so the spin-orbit
interaction is included by a variational method\cite{macdonald80}.
The spin-orbit energy is essentially converged if the
second-variational basis uses 30 LAPW eigenstates in the variational
calculation.

Energy-volume curves were calculated for mercury in the $fcc$, $bcc$
and $sc$ cubic lattices, the hexagonal close packed ($hcp$) lattice,
and the lattices described by (\ref{equ:rhom}) and (\ref{equ:bct})
above.  For convenience the latter structures will be referred to as
$\alpha$Hg and $\beta$Hg, even when they are outside the range of
parameters which properly describe these structures.  While the
cubic structures require only knowledge of the volume to determine
the structural energy, the $hcp$, $\alpha$Hg, and $\beta$Hg
structures, require a knowledge of the energy as a function of the
other lattice parameter.  For the $hcp$ and $\beta$Hg phases this
parameter is $c/a$, while for $\alpha$Hg it is the angle $\alpha$
described in (\ref{equ:rhom}).  Calculations were performed in both
the scalar-relativistic approximation and with the variational
spin-orbit energy included.

\begin{figure}[htb]
\centerline{\psfig{file=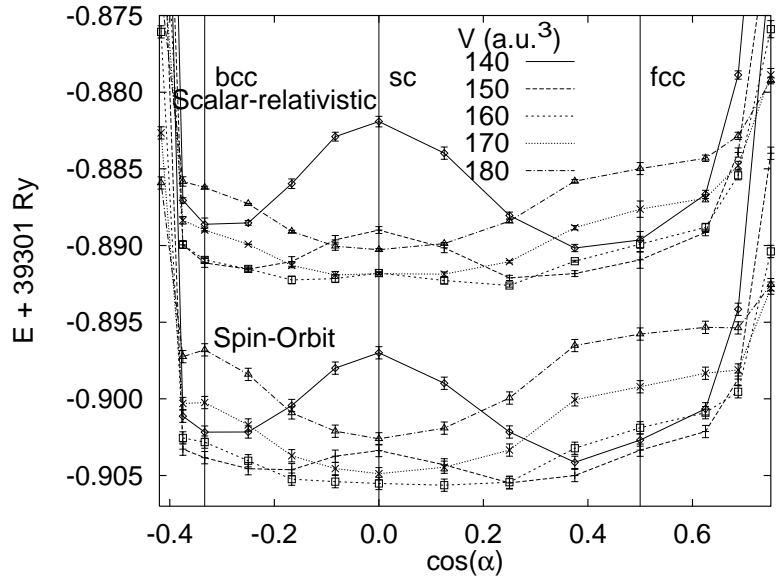,height=3.0in}}
\caption{LAPW calculation of the total energy of $\alpha$Hg in the
rhombohedral structure (system\protect{\ref{equ:rhom}}).  Both
scalar-relativistic and spin-orbit results are shown.  The error
bars represent the estimated uncertainties in the energies.  The
lines between the points are drawn as an aid to the eye.}
\label{fig:alphae}
\end{figure}

The computations for $\alpha$Hg are shown in Figure~\ref{fig:alphae}.
At most volumes there are two minima in this plot, the global
minimum between the $sc$ and $fcc$ structures, corresponding to the
observed $\alpha$Hg phase, and a secondary minimum between the $bcc$
and $sc$ structures, which will be denoted $\alpha'$Hg.  These
minima coalesce to the $sc$ structure at large volumes.  There are
no minima in the region of the $\gamma$Hg phase\cite{donohue74}.
Since this phase was obtained by shearing $\alpha$Hg, it is possible
that the phase is stabilized by a non-hydrostatic shear.

\begin{figure}[htb]
\centerline{\psfig{file=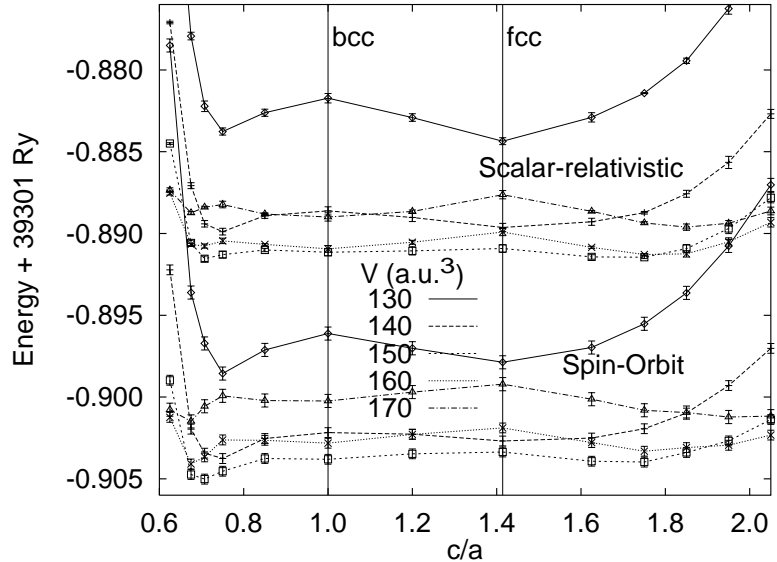,height=3.0in}}
\caption{LAPW calculation of the total energy of $\beta$Hg in the
$bct$ structure (system\protect{\ref{equ:bct}}).  Both
scalar-relativistic and spin-orbit results are shown.  The error
bars represent the estimated uncertainties in the energies.  The
lines between the points are drawn as an aid to the eye.}
\label{fig:betae}
\end{figure}

Similar calculations for $\beta$Hg are presented in
Figure~\ref{fig:betae}.
Again, there are two minima.  The first, corresponding to the
observed $\beta$Hg structure, is near $c/a \approx 0.7$.  The
second, denoted $\beta'$Hg, has a rather large shear of $c/a \approx
1.7$.  This is analogous to the large $c/a$ ratio found in the $hcp$
structures of Zn and Cd, and to the $c/a$ ratio found in the
calculations for $hcp$ Hg.  Note that without the spin-orbit
interaction the $\beta'$Hg phase is lower in energy than the
$\beta$Hg phase, but the energy difference is not significant.  The
spin-orbit interaction lowers the energy of the $\beta$Hg phase so
that it is favored over the $\beta'$Hg phase.

\begin{figure}[htb]
\centerline{\psfig{file=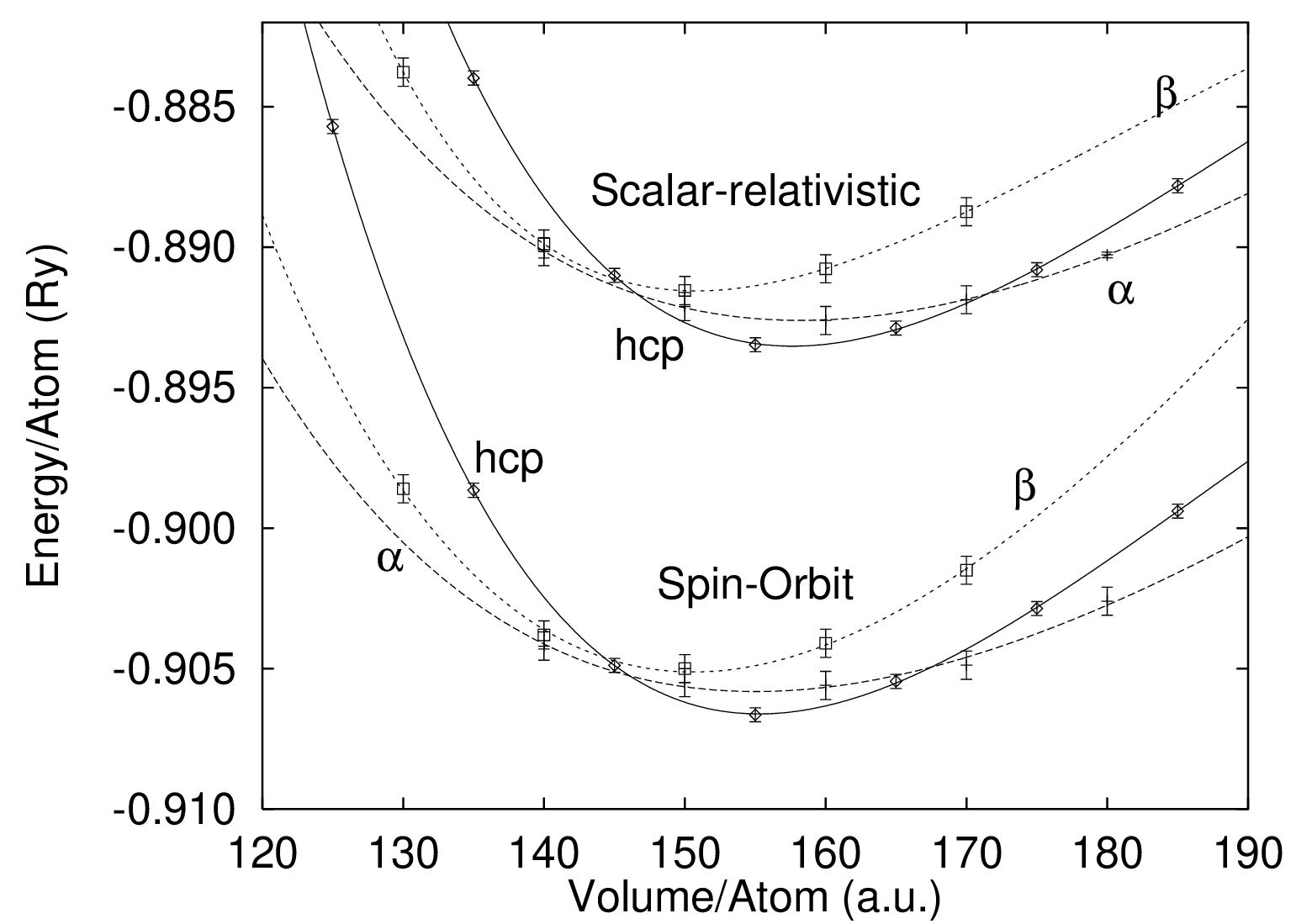,height=3.0in}}
\caption{LAPW calculation of the total energy Hg as a function of volume
for the $\alpha$Hg, $\beta$Hg, and $hcp$ phases.  Both
scalar-relativistic and spin-orbit results are shown.  The error bars
represent the estimated uncertainties in the energies.  The lines
between the points represent a Birch fit to the points shown.}
\label{fig:eos}
\end{figure}

The energy-volume curves for the low energy structures of mercury
are calculated by finding the minimum energy as a function of the
strain lattice parameter at each volume for the $\alpha$Hg,
$\beta$Hg, and $hcp$ Hg structures.  The equilibrium energies for
each of these phases is shown in Table~\ref{tab:eng}, and the full
energy volume curves are shown in Figure~\ref{fig:eos}.
\begin{table}[b]
\centering
\caption{The equilibrium energies for several structures of mercury,
shifted so that the energy of the scalar-relativistic $hcp$ phase is
set to zero.  The second column shows the scalar-relativistic
energy, the third the energy including the spin-orbit interaction,
and the fourth the difference between the two.  The ``primed''
phases are defined in the text.  The energy of atomic Hg is also
shown.}
\begin{tabular}{lrrr|lrrr}
Phase & Scalar & Spin-Orbit & \multicolumn{1}{c}{$\Delta$} &
Phase & Scalar & Spin-Orbit & \multicolumn{1}{c}{$\Delta$} \\
\hline
$fcc$ & .00244 & -.01013 & .01257 & $\alpha$ & .00092 & -.01230 &
.01322 \\
$bcc$ & .00219 & -.01010 & .01229 & $\alpha'$ & .00131 & -.01237 &
.01368 \\
$sc$ & .00139 & -.01200 & .01339 & $\beta$ & .00196 & -.01160 &
.01356 \\
$hcp$ & .00000 & -.01309 & .01309 & $\beta'$ & .00193 & -.01096 &
.01289 \\
atom & .02947 & -.00466 & .02947 & & & &
\end{tabular}
\label{tab:eng}
\end{table}
{}From the calculations we must conclude that the $hcp$ structure is
the ground state of mercury, contrary to experiment.  The spin-orbit
interaction does not change the relative ordering of the phases, and
differences in the value of the spin-orbit interaction (the $\Delta$
column in Table~\ref{tab:eng}) are numerically insignificant.  In
addition, we see that the total energy calculations cannot
distinguish between the $\alpha$Hg and the $\alpha'$Hg phases, nor
between the $\beta$Hg and the $\beta'$Hg phases.

If we nevertheless restrict ourselves to looking at the
experimentally observed phases, we find the structural properties
are in good agreement with experiment (Table~\ref{tab:struc}).
\begin{table}[b]
\centering
\caption{Structural properties of the $\alpha$Hg and $\beta$Hg phases
obtained from the LAPW calculations described in the text, compared
to experiment.}
\begin{tabular}{lrrr|rrrr}
\multicolumn{4}{c}{$\alpha$Hg} & \multicolumn{4}{c}{$\beta$Hg} \\
\hline
  & V (a.u.$^3$) & a (a.u.) & $\alpha$ (deg) & V (a.u.$^3$) & a
(a.u.) & c (a.u.) & c/a \\
\hline
Experiment & 179.7 & 5.643 & 70.743 & 152.1 & 7.549 & 5.338 &
0.7071 \\
Scalar-relativistic & 158.2 & 5.408 & 75.6 & 150.8 & 7.52 & 5.34 &
0.71 \\
Spin-Orbit & 155.0 & 5.372 & 80.2 & 150.6 & 7.52 & 5.33 & 0.71
\end{tabular}
\label{tab:struc}
\end{table}
The calculated volume of the $\alpha$Hg phase is about 13\% smaller
than the experimental volume, while the $\beta$Hg phase is only 1\%
smaller than the experimental volume.  These results are consistent
with LDA calculations for similar structures.  The calculations do
overestimate the angle ($\alpha$) for the $\alpha$Hg phase, but get
the $c/a$ ratio correctly in the $\beta$Hg phase.

\section*{\normalsize\bf DISCUSSION}
\label{sec:discuss}
{~}\\[-0.5in]

Given the many successes of the DFT, and LDA in particular, in
determining the structural properties of crystals, it is somewhat
disturbing that we cannot correctly predict the ordering of the
low-lying energy states of mercury.  There are, of course, possible
improvements to the LDA\cite{perdew92}, and these may provide part
of the answer.  A more obvious problem with the present calculation
is the form of the spin-orbit calculation.  The present
method\cite{singh94,macdonald80} uses scalar-relativistic orbitals
as a basis for the second diagonalization of the Hamiltonian
including the spin-orbit interaction.  This is a good basis set for
most states, but it has serious difficulties in dealing with the $p$
states, since the relativistic $p_{1/2}$ state is non-zero at the
origin, while the scalar-relativistic $p$ state vanishes there.
This leads to an underestimation of the spin-orbit interaction of
about 50\% in this case, as can be seen in Table~\ref{tab:eng}.  In
the atomic calculation, where the it is
calculated exactly, the spin-orbit interaction contributes 29.5~mRy
to the total energy.  For the bulk structures, however, where the
spin-orbit interaction is only approximated, it contributes about
13~mRy/atom to the total energy.  It seems likely that a better
treatment of the spin-orbit interaction will increase this
contribution.  It has been suggested\cite{singh94} that inclusion of
$p_{1/2}$-like local orbitals in the LAPW basis would improve the
spin-orbit energy, but this suggestion has yet to be implemented.

\ackno

Partial support for this work was provided by the US Office of
Naval Research.  I also wish to thank David Singh for useful
discussions concerning the spin-orbit corrections, and Larry Boyer
for discussions concerning the ``magic strains''.

\newpage

\end{document}